\documentclass[traditabstract, twocolumns]{aa}
\usepackage{graphicx}
\usepackage{txfonts}
\usepackage{natbib}
\usepackage{xspace}
\usepackage{xcolor}
\usepackage{scalerel}
\usepackage[normalem]{ulem}
\usepackage{enumitem} 
\usepackage[colorlinks=true,linkcolor=blue,citecolor=blue, urlcolor=blue]{hyperref}
\usepackage{soul}   

\usepackage{tikz}

\begin{document}

   \title{LTDE: The Lens Time Delay Experiment}
   \subtitle{I. From pixels to light curves}

   \author{Favio Neira \inst{\ref{epfl}}
          \thanks{email: \href{mailto:fcneirad@gmail.com}{fcneirad@gmail.com}}
          \and
          Frédéric Courbin \inst{\ref{barca}, \ref{icrea}}
          \and
          Frédéric Dux \inst{\ref{eso}, \ref{epfl}}
          \thanks{Disclaimer: this co-author will participate in the challenge, as such his role here is only for testing the simulation.}
          \and
          Georgios Vernardos \inst{\ref{cuny1}, \ref{cuny2}, \ref{cuny3}}}

\institute{Institute of Physics, Laboratory of Astrophysics, Ecole Polytechnique F\'ed\'erale de Lausanne (EPFL), Observatoire de Sauverny, 1290 Versoix, Switzerland \label{epfl}
\goodbreak \and
Institut de Ciències del Cosmos, Universitat de Barcelona, Martí i Franquès, 1, E-08028 Barcelona, Spain \label{barca}  
\goodbreak \and
ICREA, Pg. Llu\'is Companys 23, Barcelona, E-08010, Spain \label{icrea}
\goodbreak \and
European Southern Observatory, Alonso de Córdova 3107, Vitacura, Santiago, Chile \label{eso} 
\goodbreak \and
The Graduate Center of the City University of New York, 365 Fifth Avenue, New York, NY 10016, USA \label{cuny1}
\goodbreak \and
Department of Astrophysics, American Museum of Natural History, Central Park West and 79th Street, NY 10024-5192, USA \label{cuny2}
\goodbreak \and
Department of Physics and Astronomy, Lehman College of the CUNY, Bronx, NY 10468, USA \label{cuny3}
             }

\date{\today}

 
  \abstract {
  Gravitationally lensed quasars offer a unique opportunity to study cosmological and extragalactic phenomena, using reliable light curves of the lensed images.
  This requires accurate deblending of the quasar images, which is not trivial due to the small  separation between the lensed images (typically $\sim1$ arcsec) and because there is light contamination by the lensing galaxy and the quasar host galaxy.
   We propose a series of experiments aimed at testing our ability to extract precise and accurate photometry of lensed quasars.
   In this first paper, we focus on evaluating our ability to extract light curves from simulated CCD images of lensed quasars spanning a broad range of configurations and assuming different observational/instrumental conditions.
   Specifically, the experiment proposes to go from pixels to light curves and to evaluate the limits of current photometric algorithms.
   Our experiment has several steps, from data with known point spread function (PSF), to an unknown spatially-variable PSF field that the user has to take into account. This paper is the release of our simulated images. Anyone can extract the light curves and submit their results by the deadline. These will be evaluated with the metrics described below.
   Our set of simulations will be public and it is meant to be a benchmark for time-domain surveys like Rubin-LSST or other follow-up time-domain observations at higher temporal cadence.
   It is also meant to be a test set to help develop new algorithms in the future.
  }
  
   \keywords{Gravitational lensing -- Cosmology -- Numerical techniques}

   \maketitle

\section{Introduction}


Gravitational lensing occurs when the light of a distant source is deflected by the gravitational potential of in-between mass \citep[lens,][]{Saha2024}.
If the observer, lens and source are sufficiently aligned, the deflection of the light can be such that it will travel through several paths, producing multiple images of the same source.
In addition to this, the observed (or lensed) images can get (de)magnified.
This is due to how lensing conserves surface brightness and how the finite size of the source which can (shrink) expand from differential deflection, thus producing a (de)magnificatoin of the source.
Furthermore, if the source brightness varies, it is observed between the multiple images with a time delay, a consequence of the different length paths and gravitational potential that the light travels through \citep{Cooke1975}.

It is of particular interest when the source is a quasar lensed by a foreground galaxy, as it offers a broad range of applications \citep{Schneider1992}.
The most known one is time delay cosmography \citep{Refsdal1964}, which is an independent cosmological probe to measure the Hubble constant ($H_0$).
It can contribute to resolving the so-called Hubble tension which is the discrepancy between $H_0$ values measured from the Cosmic Microwave Background (CMB) and measurements at low redshifts \citep[e.g.][]{Verde2023}.
In addition, extracting the microlensing information from the multi-band light curves can potentially provide information on the inner structure of the lensed sources, on the microlensing objects in the lensing galaxy, and on the galaxy-scale substructures in the lensing galaxy or along its line of sight \citep{Vegetti2023}.

\begin{figure}
    \centering
    \includegraphics[width=0.45\linewidth]{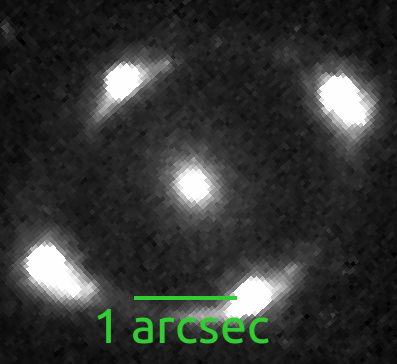}
    \includegraphics[width=0.45\linewidth]{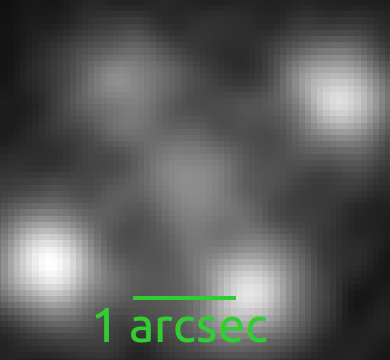}
    \includegraphics[width=0.45\linewidth]{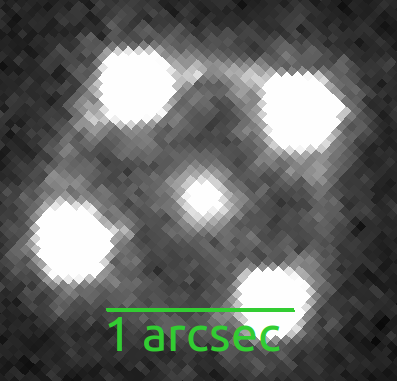}
    \includegraphics[width=0.45\linewidth]{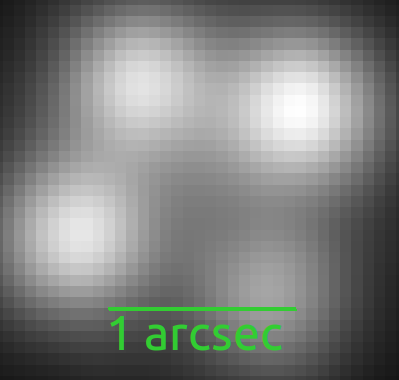}
    \caption{Top: a large image separation lensed quasar J1537-3010 \citep{lemon2019}, the left panel shows an image taken with the Hubble space telescope, in the right one taken with a ground-based telescopes. Bottom: same as top but for the lensed quasar DESJ0405-3308 \citep{Anguita2018}.}
    \label{fig:lens_examples}
\end{figure}

In order to enable these science applications, the first step is to obtain precise and accurate time delay measurements from photometric monitoring with ground-based telescopes \citep[e.g.][]{Courbin2005}.
Because the typical separation between the lensed quasar images is of the order of $\sim1$ arcsec, their photometric measurements are not trivial.
The flux from a quasar image can be contaminated by another quasar image, the host and/or the lensing galaxy, which, if not properly taken care of, can affect the accuracy of the recovered brightness variations over time and subsequently of the time delay \citep[see e.g.][]{Tewes2013,Millon2020}.
See e.g. Fig. \ref{fig:lens_examples} where we show a small and large image separation systems.
Furthermore, the number of lensed quasars increases with smaller image separation, making the deblending of quasar images more difficult.
This is of paramount importance for constraining $H_0$ for example, where the time delays of a large number of systems would be ideal \citep[see e.g.][]{Wong2020}.

To obtain photometric measurements there are several methods one can choose from, each with their own strengths and limitations.
Aperture photometry consists on measuring the brightness of the source in an aperture, then subtracting the background estimated from a surrounding annulus. Thus, it is a technique that works well for isolated bright sources but is sensitive to background noise and contamination from nearby objects.
As such, its use is not ideal for lensed quasars.
Point spread function (PSF) photometry, which models the broadening of a point source due to the atmosphere and telescope optics, can work well for blended objects \citep{Stetson1987,Bertin1996}.
Some PSF methods rely on performing a deconvolution in order to improve the spatial resolution of the image.
A few examples of these also include a decomposition of the deconvolved image in two channels, one for point sources and one for extended sources \citep[e.g.][]{Lucy1994,Magain1998,Michalewicz2023}.
Others instead rely on the use of neural networks \citep[e.g.][]{Sureau2020,Akhaury2025}.
Another photometry method is difference image analysis (DIA), which consists on comparing images taken at different times to detect and measure variable or transient sources.
However this method can be affected by alignments and changes in observing conditions across time.
Indeed several photometric surveys rely on DIA photometry \citep[e.g.][]{Kaiser2002,Udalski2003}.


In the coming years, telescopes like Euclid \citep{Euclid2024} and the Vera Rubin \citep[LSST,][]{Ivezic2019} are expected to find thousands of new gravitationally lensed quasars \citep{Oguri2010,Acevedo-Barroso2024}.
Moreover, the LSST will provide multiepoch observations in 6 bands (u, g, r, i, z, and y) for all of these systems.
Indeed, \cite{Taak2023} predicts $\sim 1000$ systems to have variability detectable with the LSST.
This is because brighter quasars show less variability \citep{Suberlak2021}.
Therefore, it will be of paramount importance to obtain accurate and precise photometry measurements that could enable cosmography and microlensing studies.
For this, one would need to quantify the performance of different photometry methods for obtaining the flux (and its error) on typical lenses with varying brightness contrasts and blendedness, as well as explore any systematic bias that may arise.

In this paper we present a challenge to the community that aims to measure our ability to do photometry of blended sources, specifically of lensed quasars.
The challenge provides several simulated cutouts of lenses, together with neighboring stars, where the participants are asked to obtain photometry of the quasar images.
The performance of the photometry method will then be quantified by metrics that are defined in this paper.
The results of this challenge will be presented in a follow up paper.

\section{Experimental setup}

The main goal of this paper is to test the performance of different photometry methods.
Specifically, when performed on strongly lensed quasars that will be observed by the LSST.
The primary sources of error that can influence the accuracy and precision of these methods are the brightness contrast between the source, host and lens galaxy, and how blended are the quasar images.
Therefore, to generate the simulated observations that will be used to test the methods, we will ignore aspects that would not affect their performance like brightness variations due to microlensing or mili-lensing by galaxy-scale substructures.
Instead, we focus on generating systems with varying lensing configurations, brightness contrasts and realistic observing conditions.
This choice was made so that the experiment is simple enough to encourage participation but realistic enough to have feasible estimates on the performance of the methods to be used on LSST data.
In the following we describe the different components used to setup the experiment.

\subsection{Simulating strongly lensed quasars}

\begin{figure*}
    \includegraphics[width=0.95\textwidth]{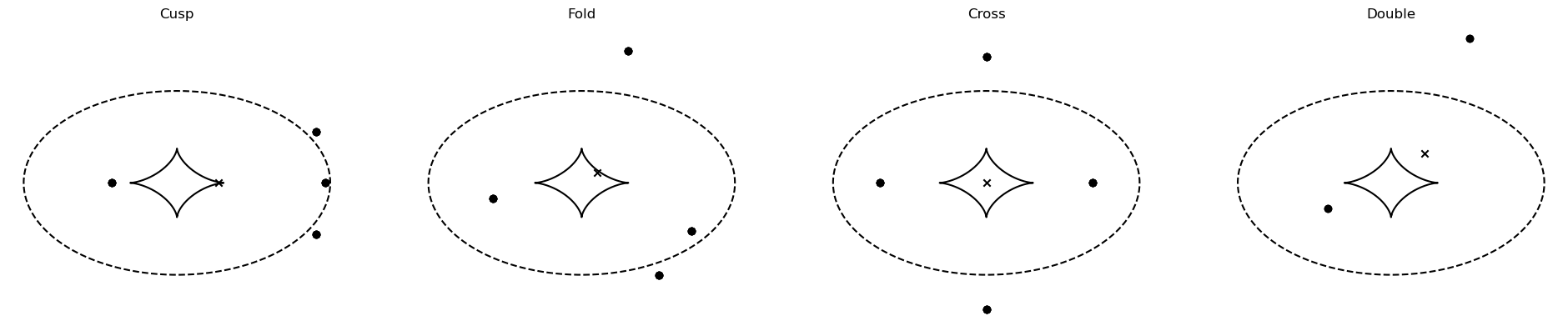}
    \caption{Examples of lens configurations. The inner (solid) and outer (dashed) caustics are plotted. The position of the quasar is denoted by the x, and the multiple images generated by a circle.}
    \label{fig:configs}
\end{figure*}

To generate the mock lensed quasars we use the MOLET software package \citep{Vernardos2022}.
In addition, we require assumptions about the mass distribution of the lensing galaxy, the light profile for the host and lens galaxy, and the intrinsic brightness of the quasar.
For simplicity, the brightness profile of both the lensing and host galaxy we adopt a Sersic profile \citep{Sersic1968}.
For the mass distribution of the lensing galaxy we adopt a singular isothermal ellipsoid, which has a projected surface density given by:
\begin{equation}
    \kappa(x, y) = \frac{1}{2} \left(\frac{\theta_{E}}{\sqrt{q x^2 + y^2/q}} \right) \,,
\end{equation}
where $\theta_{E}$ is the Einstein radius, $q$ is the minor to major axis ratio and $x$ and $y$ are the coordinates where the $x$-axis is aligned with the major axis.
The position of the multiple images will depend on the position of the source with respect to the lens caustics (formally, points in the source plane that have infinite magnification).
In this regard, there are four common ``lens configurations'', in three of which the source is quadruply imaged: ``cusps'', ``folds'' and ``crosses'', and one where the source is doubly imaged: ``double''.
We show these configurations schematically in Fig.~\ref{fig:configs}.

\subsection{Source variation}

\begin{figure}
    \includegraphics[width=0.95\linewidth]{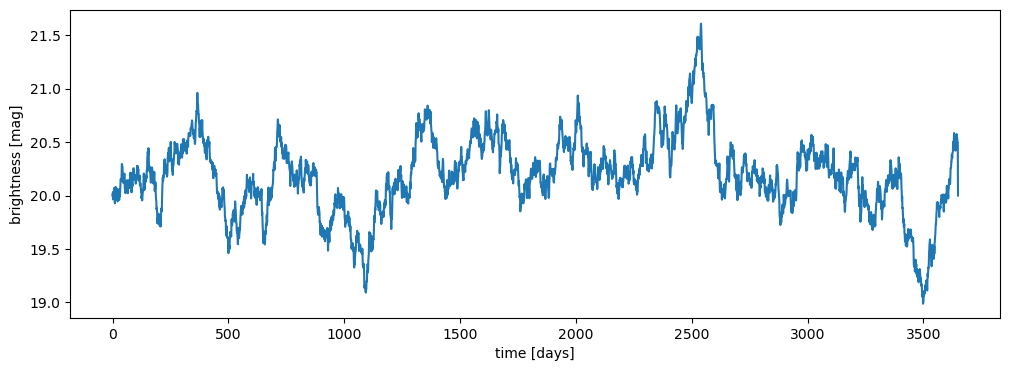}
    \caption{Example of the brightness variations of a mock quasar following a damped random walk model.}
    \label{fig:DRW}
\end{figure}

For the quasar brightness variations we assume that it follows a damped random walk (DRW), which is known to describe well typical quasar variability \citep{MacLeod2010}.
Because the goal of this paper is to quantify how well different methods perform at recovering the true quasar light curves, we can safely ignore any other sources of variability (e.g. microlenisng, reverberation, etc.).
We show an example of the variability in Fig.~\ref{fig:DRW}

\subsection{Instrumental and atmospheric effects}

To simulate the observing conditions we make use of the Vera Rubin Observatory \citep[LSST][]{Ivezic2019} data preview 0.2 (DP0.2\footnote{\url{https://dp0-2.lsst.io/}}), which is a product of the Data Challenge 2 \citep[DC2][]{DESC2021} made by the LSST Dark Energy Science Collaboration (DESC).
The data available in the DP0.2 has processed the DC2 simulated images using the LSST pipelines \citep{Bosch2019}, and provides measurements of the sky brightness and atmospheric distortion of the point spread function \citep[PSF, see][for detauls about this distortion model]{Zuntz2018}.
Although the LSST pipeline also provides a measurement of the PSF, we opt to only use their distortion measurements, which we will apply to our own PSF model.
This allows us to have full control over the shape of the PSF, and thus the true brightness of sources.
We model the PSF as a Moffat distribution:
\begin{equation}
    I(x, y) = A \left(1 + \frac{\left(x - x_{0}\right)^{2} + \left(y - y_{0}\right)^{2}}{\gamma^{2}}\right)^{- \alpha} \, ,
\end{equation}
where $A$ is a normalization constant, $x_0$ and $y_0$ are the position of the maximum of the distribution, and $\alpha$ and $\gamma$ are seeing dependent parameters.
In particular, the full width half maximum (FWHM) for a Moffat distribution is
\begin{equation}
    \text{FWHM} = 2\gamma\sqrt{2^{1/\alpha} - 1} \,.
\end{equation}
\cite{Ivezic2019} quotes expected values of FWHM that follow a log-normal distribution with mean 0.65".
To closely match this distribution, we draw values of $\gamma$ and $\alpha$ from a log-normal and uniform distribution with ranges that are consistent with observations \citep[e.g.][]{Trujillo2001}
Additionally, to avoid perfect analytical fittings, we add a shapelet component to the PSF \citep[see][for details about the shapelet components of a PSF]{Zhang2023}.

\begin{figure}
    \includegraphics[width=0.95\columnwidth]{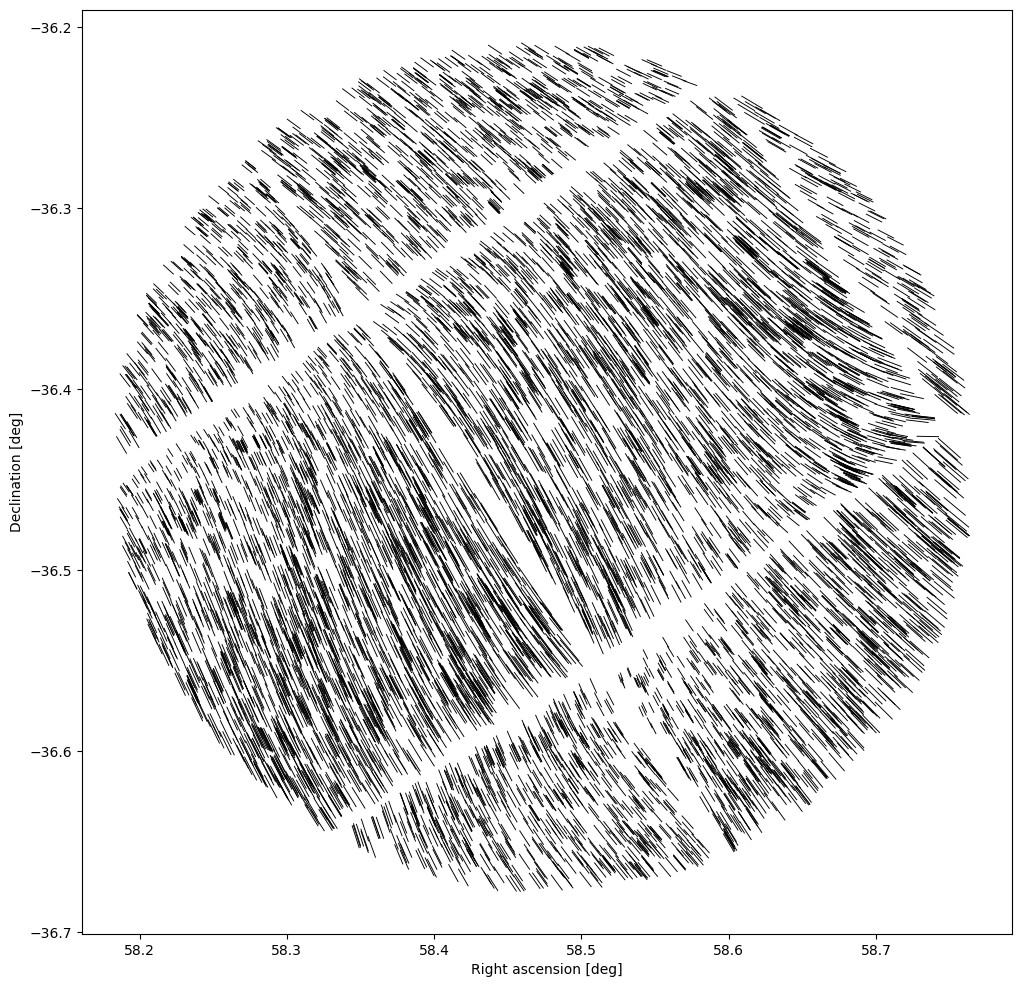}
    \caption{Example of a distortion field due to atmospheric effects taken from the LSST DP0.2. The direction of the distortion is indicated by the lines, which are exaggerated and aligned with the major axis of the ellipse that represents it.}
    \label{fig:distortion}
\end{figure}

The shape of the atmospheric PSF distortion is defined in terms of the second moments of the surface brightness profile \citep[see also][]{SeitzSchneider1997,BernsteinJarvis2002}:
\begin{equation}
    e = e_1 + ie_2 = \frac{I_{xx} - I_{yy} + 2iI_{xy}}{I_{xx} + I_{yy} + 2 \sqrt{I_{xx}I_{yy} - I_{xy}^2}} \, ,
    \label{eq:psf_e}
\end{equation}
where the moments are defined as:
\begin{equation}
    I_{\mu\nu} = \frac{\iint I(x,y) (\mu -\bar{\mu}) (\nu - \bar{\nu}) dx dy}{\iint I(x,y) dx dy} \, .
    \label{eq:psf_i}
\end{equation}
We show an example of this distortion field in Fig.~\ref{fig:distortion}.

To simulate the noise we follow the LSST documentation\footnote{See \url{https://smtn-002.lsst.io/}}.
The total instrumental noise, $\sigma^2_\text{instr}$, for each pixel is given by:
\begin{equation}
    \sigma^2_\text{instr} = (\sigma_\text{read-out}^2 + (\sigma_\text{dark} \times t_\text{exp})) \times n_\text{exp} \, ,
\end{equation}
where $\sigma_\text{read-out}=8.8$ electrons, $\sigma_\text{dark}=0.2$ electrons, $t_\text{exp}=15$ seconds and $n_\text{exp}=2$ corresponding to two back-to-back 15 seconds exposures.
Lastly, the photons from the actual astrophysical objects follow a Poissonian distribution.

\subsection{Generating LSST-like exposures}

The DP0.2 provides true (i.e. the values injected to the simulation and not the ones measured by the pipeline) photometry, astrometry and type (i.e. star or galaxy) for each source, as well as sky brightness measurements for each exposure.
The general process of generating an exposure consists of using these measurements to re-create a similar exposure.
The main motivation for re-creating instead of just injecting a lens on the exposures that are already available in the DP0.2 is, again, to have full control on the true PSF and brightness of the sources.

\begin{figure}
    \includegraphics[width=0.95\linewidth]{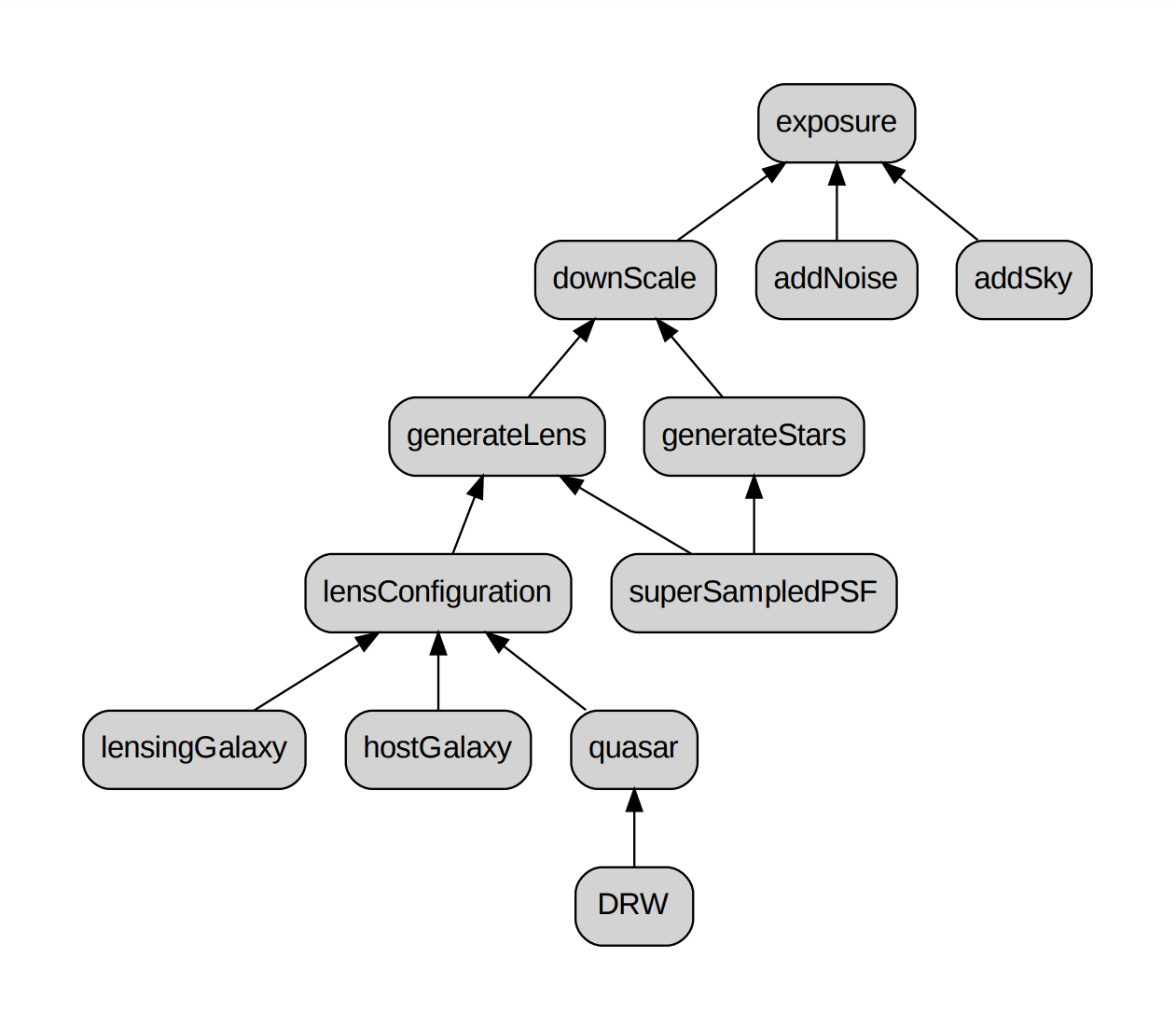}
    \caption{Diagram showing how an exposure is generated}
    \label{fig:pipeline}
\end{figure}

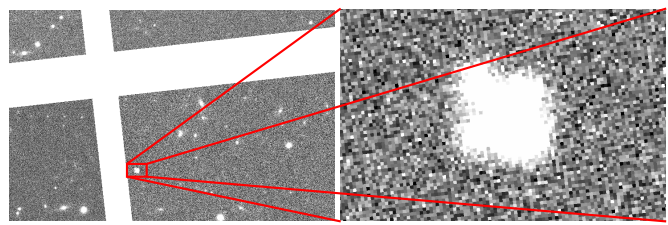
\begin{figure}
    \centering
    \begin{tikzpicture}
    \node[anchor=south west,inner sep=0] (image) at (0,0) {\includegraphics[width=0.475\columnwidth]{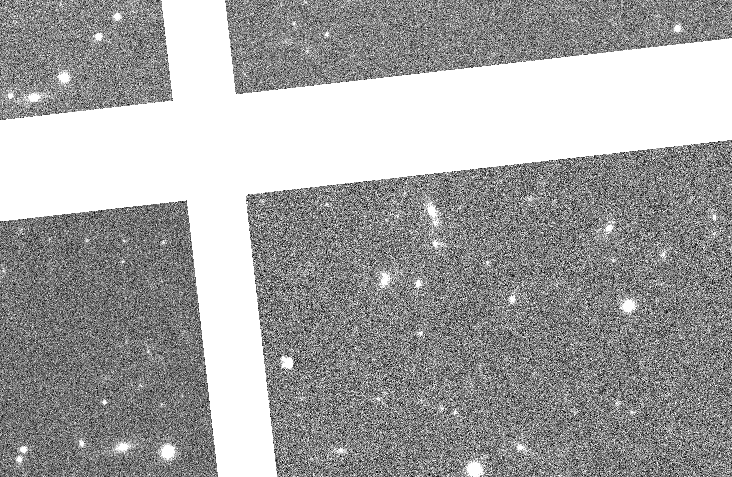} \includegraphics[width=0.475\columnwidth]{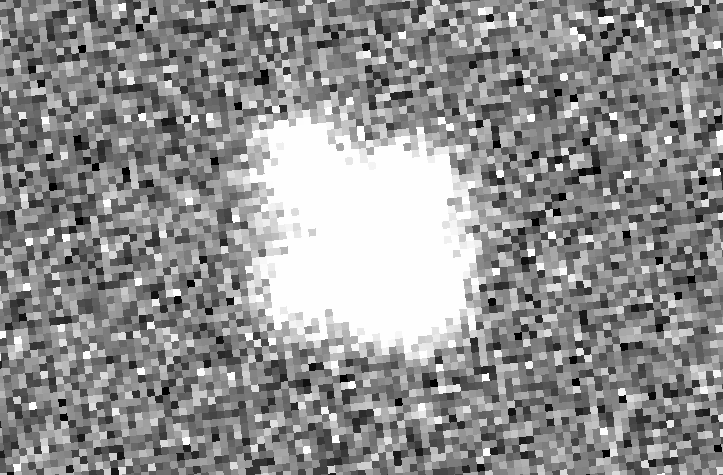}};
    \begin{scope}[x={(image.south east)},y={(image.north west)}]
    \draw[red, thick] (0.18,0.21) rectangle (0.21,0.27);
    \draw[red, thick] (0.18,0.21) -- (0.505,0);
    \draw[red, thick] (0.18,0.27) -- (0.505,1);
    \draw[red, thick] (0.21,0.21) -- (1,0);
    \draw[red, thick] (0.21,0.27) -- (1,1);
    \end{scope}
    \end{tikzpicture}
    \caption{Left: example of a simulated exposure Right: zoomed window where the lens is injected. Note the gap between the detectors which will have different orientations for each different exposure.}
    \label{fig:exposure}
\end{figure}

To get the data from DP0.2 we start by selecting a region on the sky, then query the exposures and catalogues\footnote{This is done in the Rubin science platform at \url{https://data.lsst.cloud/}}.
We end up with a catalogue that contains the flux, position, and type for each source that lies in the selected region.
The exposures that are queried come with a world coordinate system (WCS) which is re-used in order to keep the rotations and gaps between the detectors consistent with the LSST.
With this, together with the photometry and astrometry that are provided, we are able to inject the sources into the detectors.
The lensed quasar is injected at the position (and as a replacement) where a star is originally detected.
To do this we use a super-sampled PSF (10 times the LSST resolution) to convolve the source image to be injected, down-scaling to the pixel size of the LSST (0.2") and finally adding the provided sky brightness provided and a noise layer composed of instrumental plus Poissonian components.
In Fig.~\ref{fig:pipeline} we show a graph representing these steps and a resulting example of an exposure in Fig.~\ref{fig:exposure}.

\section{The mock samples}

\begin{figure*}
    \centering
    \includegraphics[width=0.95\linewidth]{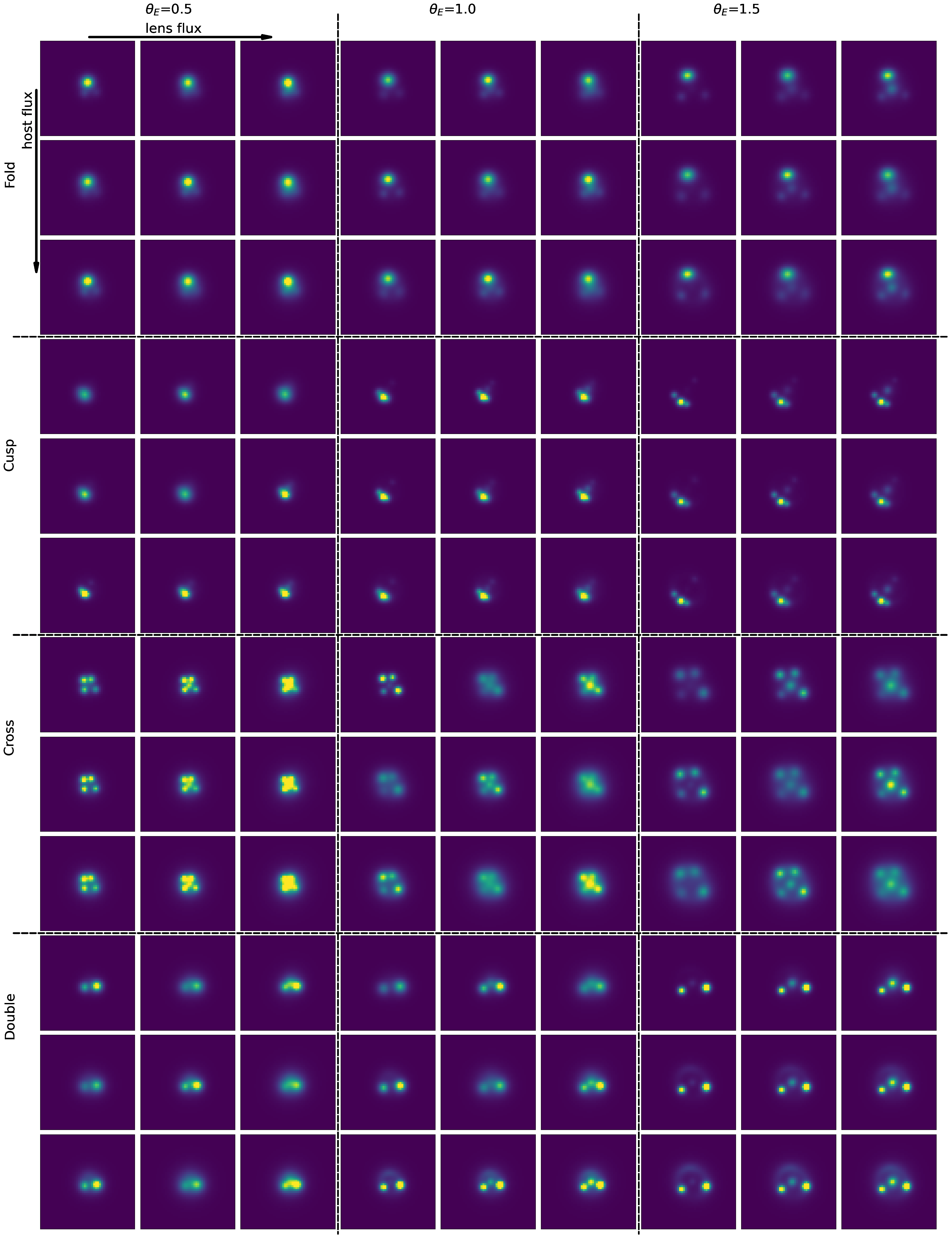}
    \caption{All 108 mock cutouts. The cutouts are divided as follows. Groups of $(3 \times 3)$ have the same $\theta_\text{E}$ and lens configuration, and vary the brightness of the host (row) and lensing (column) galaxy. The top group of $(3 \times 9)$ have the same lens configuration (fold) and every 3 columns $\theta_\text{E}$ varies. The following 3 $(3 \times 9)$ groups are the same as the one above but for a different lens configuration (cusp, cross and double).}
    \label{fig:mocks}
\end{figure*}

The mock sample consists of each lens configuration shown in Fig. \ref{fig:configs} with varying brightness contrast and $\theta_\text{E}$ scaling.
Compared to a fiducial mock system, we vary:
\begin{itemize}
    \item The host galaxy brightness by a factor of either $0.1$ and $0.5$.
    \item The lensing galaxy brightness by a factor of either $0.1$ and $0.5$.
    \item The Einstein radius by a factor of either $0.75$ and $1.5$.
\end{itemize}
This yields a total of $3 \times 3 \times 3 = 27$ mocks for a single lens configuration, thus a total of $27 \times 4 = 108$ mock systems.
We show examples of these mocks in Fig.~\ref{fig:mocks}.
The instructions to get the data are in \url{https://obswww.unige.ch/~neiradia/}.

\subsection{Round \#0}

The purpose of this round is for the participants to explore the systematic strengths and weaknesses of their method.
As such, in addition to the cutouts of the system and neighbouring stars, we provide a noiseless PSF that is constant across the field of view, but varies in time.
The noiseless PSF allows to get the most accurate light curve.
This, paired with the multiple configurations of the systems, will allow to assess the methods performance for different image blendedness and brightness contrasts.

\subsection{Round \#1}

For this round only the cutouts will be provided.
The true PSF will remain unknown, and it will be the task of the participants to obtain their own PSF.
As in the previous round, here the PSF is constant across the field of view but changes at every time stamp.

\subsection{Round \#2}

The last round introduces an atmospheric distortion (see equations \ref{eq:psf_e} and \ref{eq:psf_i}) to the PSF that changes across the field of view, we show an example of this in Fig.~\ref{fig:distortion}.
This round will allow us to assess how much this distortion can affect the accuracy of the light curves.

\section{Metrics}
\label{sec:metrics}

In order to quantify how well the different methods perform at recovering the light curves we need to define a standardized metric.
For each method, we will assess:
\begin{itemize}
    \item The flux leakage between the different components of the system.
    \item The accuracy of the photometry.
    \item The accuracy of the noise estimation.
\end{itemize}
For the flux leakage, a $\chi^2$ test between the true and submitted light curves will be performed, defined as:
\begin{equation}
    \chi^2_n = \sum_{i=1}^{N_\text{epochs}} \frac{(y_{n, i} - \overline{y_{n, i}})^2}{\sigma_{n, i}^2} \,,
\end{equation}
where $y$ and $\sigma$ are the measured brightness and its uncertainty respectively, and $\overline{y}$ stands for the true brightness.
The indices $n$ and $i$ denote the corresponding image of the system and observation epoch respectively.
Because we will look at the values of $\chi^2_n$ separately for each quasar image, this will allow us to measure how much flux from one (or more) image leaks into another.

For the accuracy of the photometry, we will use a $\chi^2$ test marginalised by a constant offset parameter \citep[see][]{Bretthorst1988} on the combined brightness of all quasar images, which is defined as:
\begin{align}
    &\chi^2_\text{offset} = \alpha - \frac{\beta^2}{4\gamma} \, \\
    &\alpha = \sum_{i=1}^{N_\text{epochs}} \frac{(\sum_{n=1}^{N_\text{img}}y_{n,i} - \sum_{n=1}^{N_\text{img}}\overline{y_{n,i}})^2}{\sum_{n=1}^{N_\text{img}}\sigma_{n,i}^2} \, \\
    &\beta = -2 \sum_{i=1}^{N_\text{epochs}} \frac{(\sum_{n=1}^{N_\text{img}}y_{n,i} - \sum_{n=1}^{N_\text{img}}\overline{y_{n,i}})}{\sum_{n=1}^{N_\text{img}}\sigma_{n,i}^2} \, \\
    &\gamma = \sum_{i=1}^{N_\text{epochs}} \frac{1}{\sum_{n=1}^{N_\text{img}}\sigma_{n,i}^2} \,.
\end{align}
Unlike the previous metric, by combining the brightness of the images we are effectively ignoring any leakage between them.
Furthermore, because this metric is oblivious to a constant offset in the light curve, it is useful as a measure of how well the variations over time are recovered while ignoring any offset that may come from the background and/or the lens.
Finally, to quantify how well the flux errors are recovered, we simply compute the absolute percentage error as:
\begin{equation}
    \epsilon_{n} = \sum_{i=1}^{N_\text{epochs}} \frac{|\sigma_{n,i} - \overline{\sigma_{n,i}}|}{\overline{\sigma_{n,i}}} \,,
\end{equation}
where $\overline{\sigma_{n,i}}$ is the true uncertainty defined by the sum in quadrature of instrumental and photon noise.

The metrics described here will be used to evaluate all submissions of the participants.
However, we reserve the right to modify or replace these at a later stage and informing all the participants about the final metrics that will be used.

\section{Participation to the experiment}

The challenge presented here (for advertising purposes) is part of the Dark Energy Science Collaboration (DESC).
However, it is open to everyone regardless of whether participants are members of the DESC or not.
To allow this, participants who are not members of the DESC will be asked to sign a code of conduct.

To obtain the simulated data, participants can visit the following website \url{https://obswww.unige.ch/~neiradia/}.
There, we host information on how to download and visualize the data.

Participants will be asked to submit their results in a specific format.
For each of the configurations (see Fig. \ref{fig:configs}), their respective light curves should be in a \texttt{csv} file.
This file should include the time [MJD]\footnote{This can be found in the header of each fits file.}, measured fluxes, and the flux errors at each epoch.
The naming of these variables should be included in the header of the \texttt{csv} file and should comply to the following conventions:
\begin{itemize}
    \item The time should be named \texttt{time}.
    \item The flux for each quasar image should be named \texttt{flux\_A}, \texttt{flux\_B}, \texttt{flux\_C} and \texttt{flux\_D}, where \texttt{\_A} indicates the brightest image \texttt{\_B} the second, and so on.
    \item The flux errors should be named \texttt{flux\_err\_A}, \texttt{flux\_err\_B}, \texttt{flux\_err\_C} and \texttt{flux\_err\_D}, following the same convention as the flux.
    \item If the measurement of two or more quasar images cannot be isolated, their measurement should be named using the suffixes of the quasar image (e.g. \texttt{flux\_AB}). In this case, the measurement values that are not feasible should be set to \texttt{nan}.
    \item If the measurement of any quasar image at any epoch cannot be done, the \texttt{time} should be set to its corresponding epoch, while all others set to \texttt{nan}.
\end{itemize}

\noindent The deadlines for submissions are the following:
\begin{itemize}
    \item Round \#0: training set - no deadline.
    \item Round \#1: 30 May 2025.
    \item Round \#2: 30 June 2025.
\end{itemize}
Note that deadlines may change slightly depending on participation in the experiment, and that independently of the deadlines our data can be used freely anytime to test or develop new algorithms. However, the input used to produce the simulations will not be revealed before the publication of the results with the contributions of all participating teams.

After the deadline, we will perform an analysis comparing the submissions with the true data by using the metrics described in Section \ref{sec:metrics}.
The results will be presented in a follow-up paper, in which the participants will be asked to summarize their methods.

\section{Conclusion}

In the present work, we design a simple experiment with multiple objectives. First, we propose to test existing techniques to extract time series of lensed point sources spanning a wide range of image configurations and lens and source properties. Second, the same simulations can be used to develop new techniques, benchmark them against the performances of previous techniques, and evaluate their relative merits and limitations. Finally, at the end of the present experiment, we make public the codes and tools used to produce the simulations, so that future experiments can be developed based on our work to test or isolate specific aspects or to develop a full end-to-end simulation pipeline to test photometric methods.

In the long run, we believe that the present experiment is the first of a modular pipeline or methodology to evaluate the error budget of time delay cosmography with lensed quasars and supernovae, propagating the errors all the way from the extraction of light curves from pixels, to the influence of the PSF construction itself or the contamination by astrophysical effects such as microlensing, dust or the internal structure of the lensed sources.


\begin{acknowledgements}
This is supported by the Swiss National Science Foundation (SNSF). Put here the formal acknowledgment for FutureLens. 

\end{acknowledgements}

%
%

\bibliographystyle{aa}
\bibliography{aanda}

\begin{thebibliography}{38}
\expandafter\ifx\csname natexlab\endcsname\relax\def\natexlab#1{#1}\fi

\bibitem[{{Acevedo Barroso} {et~al.}(2024){Acevedo Barroso}, {O'Riordan},
  {Cl{\'e}ment}, {Tortora}, {Collett}, {Courbin}, {Gavazzi}, {Metcalf},
  {Busillo}, {Andika}, {Cabanac}, {Courtois}, {Crook-Mansour}, {Delchambre},
  {Despali}, {Ecker}, {Franco}, {Holloway}, {Jackson}, {Jahnke}, {Mahler},
  {Marchetti}, {Matavulj}, {Melo}, {Meneghetti}, {Moustakas}, {M{\"u}ller},
  {Nucita}, {Paulino-Afonso}, {Pearson}, {Rojas}, {Scarlata}, {Schuldt},
  {Serjeant}, {Sluse}, {Suyu}, {Vaccari}, {Verma}, {Vernardos}, {Walmsley},
  {Bouy}, {Walth}, {Powell}, {Bolzonella}, {Cuillandre}, {Kluge}, {Saifollahi},
  {Schirmer}, {Stone}, {Acebron}, {Bazzanini}, {D{\'\i}az-S{\'a}nchez}, {Hogg},
  {Koopmans}, {Kruk}, {Leuzzi}, {Manj{\'o}n-Garc{\'\i}a}, {Mannucci}, {Nagam},
  {Pearce-Casey}, {Scharr{\'e}}, {Wilde}, {Altieri}, {Amara}, {Andreon},
  {Auricchio}, {Baccigalupi}, {Baldi}, {Balestra}, {Bardelli}, {Basset},
  {Battaglia}, {Bender}, {Bonino}, {Branchini}, {Brescia}, {Brinchmann},
  {Caillat}, {Camera}, {Candini}, {Capobianco}, {Carbone}, {Carretero},
  {Casas}, {Castellano}, {Castignani}, {Cavuoti}, {Cimatti}, {Colodro-Conde},
  {Congedo}, {Conselice}, {Conversi}, {Copin}, {Corcione}, {Cropper}, {Da
  Silva}, {Degaudenzi}, {De Lucia}, {Dinis}, {Dubath}, {Dupac}, {Dusini},
  {Farina}, {Farrens}, {Ferriol}, {Frailis}, {Franceschi}, {Galeotta},
  {Garilli}, {George}, {Gillard}, {Gillis}, {Giocoli}, {G{\'o}mez-Alvarez},
  {Grazian}, {Grupp}, {Guzzo}, {Haugan}, {Hoekstra}, {Holmes}, {Hook},
  {Hormuth}, {Hornstrup}, {Jhabvala}, {Joachimi}, {Keih{\"a}nen}, {Kermiche},
  {Kiessling}, {Kubik}, {Kunz}, {Kurki-Suonio}, {Le Mignant}, {Ligori},
  {Lilje}, {Lindholm}, {Lloro}, {Mainetti}, {Maiorano}, {Mansutti}, {Marcin},
  {Marggraf}, {Martinelli}, {Martinet}, {Marulli}, {Massey}, {Medinaceli},
  {Melchior}, {Mellier}, {Merlin}, {Meylan}, {Moresco}, {Moscardini}, {Munari},
  {Nakajima}, {Neissner}, {Nichol}, {Niemi}, {Nightingale}, {Padilla},
  {Paltani}, {Pasian}, {Pedersen}, {Percival}, {Pettorino}, {Pires}, {Polenta},
  {Poncet}, {Popa}, {Pozzetti}, {Raison}, {Rebolo}, {Renzi}, {Rhodes},
  {Riccio}, {Romelli}, {Roncarelli}, {Rossetti}, {Saglia}, {Sakr},
  {S{\'a}nchez}, {Sapone}, {Schneider}, {Schrabback}, {Secroun}, {Seidel},
  {Serrano}, {Sirignano}, {Sirri}, {Skottfelt}, {Stanco}, {Steinwagner},
  {Tallada-Cresp{\'\i}}, {Tavagnacco}, {Taylor}, {Tereno}, {Toledo-Moreo},
  {Torradeflot}, {Tutusaus}, {Valentijn}, \&
  {Valenziano}}]{Acevedo-Barroso2024}
{Acevedo Barroso}, J.~A., {O'Riordan}, C.~M., {Cl{\'e}ment}, B., {et~al.} 2024,
  arXiv e-prints, arXiv:2408.06217

\bibitem[{{Akhaury} {et~al.}(2025){Akhaury}, {Jablonka}, {Courbin}, \&
  {Starck}}]{Akhaury2025}
{Akhaury}, U., {Jablonka}, P., {Courbin}, F., \& {Starck}, J.-L. 2025, arXiv
  e-prints, arXiv:2502.17177

\bibitem[{{Anguita} {et~al.}(2018){Anguita}, {Schechter}, {Kuropatkin},
  {Morgan}, {Ostrovski}, {Abramson}, {Agnello}, {Apostolovski}, {Fassnacht},
  {Hsueh}, {Motta}, {Rojas}, {Rusu}, {Treu}, {Williams}, {Auger},
  {Buckley-Geer}, {Lin}, {McMahon}, {Abbott}, {Allam}, {Annis}, {Bernstein},
  {Bertin}, {Brooks}, {Burke}, {Carnero Rosell}, {Carrasco-Kind}, {Carretero},
  {Cunha}, {D'Andrea}, {De Vicente}, {DePoy}, {Desai}, {Diehl}, {Doel},
  {Flaugher}, {Garc{\'\i}a-Bellido}, {Gerdes}, {Gruen}, {Gruendl}, {Gschwend},
  {Hartley}, {Hollowood}, {Honscheid}, {James}, {Kuehn}, {Lima}, {Maia},
  {Miquel}, {Plazas}, {Sanchez}, {Scarpine}, {Smith}, {Soares-Santos},
  {Sobreira}, {Suchyta}, {Tarle}, \& {Walker}}]{Anguita2018}
{Anguita}, T., {Schechter}, P.~L., {Kuropatkin}, N., {et~al.} 2018, \mnras,
  480, 5017

\bibitem[{{Bernstein} \& {Jarvis}(2002)}]{BernsteinJarvis2002}
{Bernstein}, G.~M. \& {Jarvis}, M. 2002, \aj, 123, 583

\bibitem[{{Bertin} \& {Arnouts}(1996)}]{Bertin1996}
{Bertin}, E. \& {Arnouts}, S. 1996, \aaps, 117, 393

\bibitem[{{Bosch} {et~al.}(2019){Bosch}, {AlSayyad}, {Armstrong}, {Bellm},
  {Chiang}, {Eggl}, {Findeisen}, {Fisher-Levine}, {Guy}, {Guyonnet},
  {Ivezi{\'c}}, {Jenness}, {Kov{\'a}cs}, {Krughoff}, {Lupton}, {Lust},
  {MacArthur}, {Meyers}, {Moolekamp}, {Morrison}, {Morton}, {O'Mullane},
  {Parejko}, {Plazas}, {Price}, {Rawls}, {Reed}, {Schellart}, {Slater},
  {Sullivan}, {Swinbank}, {Taranu}, {Waters}, \& {Wood-Vasey}}]{Bosch2019}
{Bosch}, J., {AlSayyad}, Y., {Armstrong}, R., {et~al.} 2019, in Astronomical
  Society of the Pacific Conference Series, Vol. 523, Astronomical Data
  Analysis Software and Systems XXVII, ed. P.~J. {Teuben}, M.~W. {Pound}, B.~A.
  {Thomas}, \& E.~M. {Warner}, 521

\bibitem[{Bretthorst(1988)}]{Bretthorst1988}
Bretthorst, G.~L. 1988, Bayesian spectrum Analysis and parameter estimation
  (Springer-Verlag Berlin Heidelberg)

\bibitem[{{Cooke} \& {Kantowski}(1975)}]{Cooke1975}
{Cooke}, J.~H. \& {Kantowski}, R. 1975, \apjl, 195, L11

\bibitem[{{Courbin} {et~al.}(2005){Courbin}, {Eigenbrod}, {Vuissoz}, {Meylan},
  \& {Magain}}]{Courbin2005}
{Courbin}, F., {Eigenbrod}, A., {Vuissoz}, C., {Meylan}, G., \& {Magain}, P.
  2005, in IAU Symposium, Vol. 225, Gravitational Lensing Impact on Cosmology,
  ed. Y.~{Mellier} \& G.~{Meylan}, 297--303

\bibitem[{{Euclid Collaboration} {et~al.}(2024){Euclid Collaboration},
  {Mellier}, {Abdurro'uf}, {Acevedo Barroso}, {Ach{\'u}carro}, {Adamek},
  {Adam}, {Addison}, {Aghanim}, {Aguena}, {Ajani}, {Akrami}, {Al-Bahlawan},
  {Alavi}, {Albuquerque}, {Alestas}, {Alguero}, {Allaoui}, {Allen}, {Allevato},
  {Alonso-Tetilla}, {Altieri}, {Alvarez-Candal}, {Alvi}, {Amara}, {Amendola},
  {Amiaux}, {Andika}, {Andreon}, {Andrews}, {Angora}, {Angulo}, {Annibali},
  {Anselmi}, {Anselmi}, {Arcari}, {Archidiacono}, {Aric{\`o}}, {Arnaud},
  {Arnouts}, {Asgari}, {Asorey}, {Atayde}, {Atek}, {Atrio-Barandela}, {Aubert},
  {Aubourg}, {Auphan}, {Auricchio}, {Aussel}, {Aussel}, {Avelino},
  {Avgoustidis}, {Avila}, {Awan}, {Azzollini}, {Baccigalupi}, {Bachelet},
  {Bacon}, {Baes}, {Bagley}, {Bahr-Kalus}, {Balaguera-Antolinez}, {Balbinot},
  {Balcells}, {Baldi}, {Baldry}, {Balestra}, {Ballardini}, {Ballester},
  {Balogh}, {Ba{\~n}ados}, {Barbier}, {Bardelli}, {Baron}, {Barreiro},
  {Barrena}, {Barriere}, {Barros}, {Barthelemy}, {Bartolo}, {Basset},
  {Battaglia}, {Battisti}, {Baugh}, {Baumont}, {Bazzanini}, {Beaulieu},
  {Beckmann}, {Belikov}, {Bel}, {Bellagamba}, {Bella}, {Bellini}, {Benabed},
  {Bender}, {Benevento}, {Bennett}, {Benson}, {Bergamini}, {Bermejo-Climent},
  {Bernardeau}, {Bertacca}, {Berthe}, {Berthier}, {Bethermin}, {Beutler},
  {Bevillon}, {Bhargava}, {Bhatawdekar}, {Bianchi}, {Bisigello}, {Biviano},
  {Blake}, {Blanchard}, {Blazek}, {Blot}, {Bosco}, {Bodendorf}, {Boenke},
  {B{\"o}hringer}, {Boldrini}, {Bolzonella}, {Bonchi}, {Bonici}, {Bonino},
  {Bonino}, {Bonvin}, {Bon}, {Booth}, {Borgani}, {Borlaff}, {Borsato}, {Bosco},
  {Bose}, {Botticella}, {Boucaud}, {Bouche}, {Boucher}, {Boutigny}, {Bouvard},
  {Bouwens}, {Bouy}, {Bowler}, {Bozza}, {Bozzo}, {Branchini}, {Brando},
  {Brau-Nogue}, {Brekke}, {Bremer}, {Brescia}, {Breton}, {Brinchmann},
  {Brinckmann}, {Brockley-Blatt}, {Brodwin}, {Brouard}, {Brown}, {Bruton},
  {Bucko}, {Buddelmeijer}, {Buenadicha}, {Buitrago}, {Burger}, {Burigana},
  {Busillo}, {Busonero}, {Cabanac}, {Cabayol-Garcia}, {Cagliari}, {Caillat},
  {Caillat}, {Calabrese}, {Calabro}, {Calderone}, {Calura}, {Camacho Quevedo},
  {Camera}, {Campos}, {Canas-Herrera}, {Candini}, {Cantiello}, {Capobianco},
  {Cappellaro}, {Cappelluti}, {Cappi}, {Caputi}, {Cara}, {Carbone}, {Cardone},
  {Carella}, {Carlberg}, {Carle}, {Carminati}, {Caro}, {Carrasco}, {Carretero},
  {Carrilho}, \& {Carron Duque}}]{Euclid2024}
{Euclid Collaboration}, {Mellier}, Y., {Abdurro'uf}, {et~al.} 2024, arXiv
  e-prints, arXiv:2405.13491

\bibitem[{{Ivezi{\'c}} {et~al.}(2019){Ivezi{\'c}}, {Kahn}, {Tyson}, {Abel},
  {Acosta}, {Allsman}, {Alonso}, {AlSayyad}, {Anderson}, {Andrew}, {Angel},
  {Angeli}, {Ansari}, {Antilogus}, {Araujo}, {Armstrong}, {Arndt}, {Astier},
  {Aubourg}, {Auza}, {Axelrod}, {Bard}, {Barr}, {Barrau}, {Bartlett}, {Bauer},
  {Bauman}, {Baumont}, {Bechtol}, {Bechtol}, {Becker}, {Becla}, {Beldica},
  {Bellavia}, {Bianco}, {Biswas}, {Blanc}, {Blazek}, {Blandford}, {Bloom},
  {Bogart}, {Bond}, {Booth}, {Borgland}, {Borne}, {Bosch}, {Boutigny},
  {Brackett}, {Bradshaw}, {Brandt}, {Brown}, {Bullock}, {Burchat}, {Burke},
  {Cagnoli}, {Calabrese}, {Callahan}, {Callen}, {Carlin}, {Carlson},
  {Chandrasekharan}, {Charles-Emerson}, {Chesley}, {Cheu}, {Chiang}, {Chiang},
  {Chirino}, {Chow}, {Ciardi}, {Claver}, {Cohen-Tanugi}, {Cockrum}, {Coles},
  {Connolly}, {Cook}, {Cooray}, {Covey}, {Cribbs}, {Cui}, {Cutri}, {Daly},
  {Daniel}, {Daruich}, {Daubard}, {Daues}, {Dawson}, {Delgado}, {Dellapenna},
  {de Peyster}, {de Val-Borro}, {Digel}, {Doherty}, {Dubois},
  {Dubois-Felsmann}, {Durech}, {Economou}, {Eifler}, {Eracleous}, {Emmons},
  {Fausti Neto}, {Ferguson}, {Figueroa}, {Fisher-Levine}, {Focke}, {Foss},
  {Frank}, {Freemon}, {Gangler}, {Gawiser}, {Geary}, {Gee}, {Geha}, {Gessner},
  {Gibson}, {Gilmore}, {Glanzman}, {Glick}, {Goldina}, {Goldstein}, {Goodenow},
  {Graham}, {Gressler}, {Gris}, {Guy}, {Guyonnet}, {Haller}, {Harris},
  {Hascall}, {Haupt}, {Hernandez}, {Herrmann}, {Hileman}, {Hoblitt}, {Hodgson},
  {Hogan}, {Howard}, {Huang}, {Huffer}, {Ingraham}, {Innes}, {Jacoby}, {Jain},
  {Jammes}, {Jee}, {Jenness}, {Jernigan}, {Jevremovi{\'c}}, {Johns}, {Johnson},
  {Johnson}, {Jones}, {Juramy-Gilles}, {Juri{\'c}}, {Kalirai}, {Kallivayalil},
  {Kalmbach}, {Kantor}, {Karst}, {Kasliwal}, {Kelly}, {Kessler}, {Kinnison},
  {Kirkby}, {Knox}, {Kotov}, {Krabbendam}, {Krughoff}, {Kub{\'a}nek},
  {Kuczewski}, {Kulkarni}, {Ku}, {Kurita}, {Lage}, {Lambert}, {Lange},
  {Langton}, {Le Guillou}, {Levine}, {Liang}, {Lim}, {Lintott}, {Long},
  {Lopez}, {Lotz}, {Lupton}, {Lust}, {MacArthur}, {Mahabal}, {Mandelbaum},
  {Markiewicz}, {Marsh}, {Marshall}, {Marshall}, {May}, {McKercher}, {McQueen},
  {Meyers}, {Migliore}, {Miller}, {Mills}, {Miraval}, {Moeyens}, {Moolekamp},
  {Monet}, {Moniez}, {Monkewitz}, {Montgomery}, {Morrison}, {Mueller},
  {Muller}, {Mu{\~n}oz Arancibia}, {Neill}, {Newbry}, {Nief}, {Nomerotski},
  {Nordby}, {O'Connor}, {Oliver}, {Olivier}, {Olsen}, {O'Mullane}, {Ortiz},
  {Osier}, {Owen}, {Pain}, {Palecek}, {Parejko}, {Parsons}, {Pease},
  {Peterson}, {Peterson}, {Petravick}, {Libby Petrick}, {Petry},
  {Pierfederici}, {Pietrowicz}, {Pike}, {Pinto}, {Plante}, {Plate}, {Plutchak},
  {Price}, {Prouza}, {Radeka}, {Rajagopal}, {Rasmussen}, {Regnault}, {Reil},
  {Reiss}, {Reuter}, {Ridgway}, {Riot}, {Ritz}, {Robinson}, {Roby}, {Roodman},
  {Rosing}, {Roucelle}, {Rumore}, {Russo}, {Saha}, {Sassolas}, {Schalk},
  {Schellart}, {Schindler}, {Schmidt}, {Schneider}, {Schneider}, {Schoening},
  {Schumacher}, {Schwamb}, {Sebag}, {Selvy}, {Sembroski}, {Seppala}, {Serio},
  {Serrano}, {Shaw}, {Shipsey}, {Sick}, {Silvestri}, {Slater}, {Smith},
  {Smith}, {Sobhani}, {Soldahl}, {Storrie-Lombardi}, {Stover}, {Strauss},
  {Street}, {Stubbs}, {Sullivan}, {Sweeney}, {Swinbank}, {Szalay}, {Takacs},
  {Tether}, {Thaler}, {Thayer}, {Thomas}, {Thornton}, {Thukral}, {Tice},
  {Trilling}, {Turri}, {Van Berg}, {Vanden Berk}, {Vetter}, {Virieux},
  {Vucina}, {Wahl}, {Walkowicz}, {Walsh}, {Walter}, {Wang}, {Wang}, {Warner},
  {Wiecha}, {Willman}, {Winters}, {Wittman}, {Wolff}, {Wood-Vasey}, {Wu},
  {Xin}, {Yoachim}, \& {Zhan}}]{Ivezic2019}
{Ivezi{\'c}}, {\v{Z}}., {Kahn}, S.~M., {Tyson}, J.~A., {et~al.} 2019, \apj,
  873, 111

\bibitem[{{Kaiser} {et~al.}(2002){Kaiser}, {Aussel}, {Burke}, {Boesgaard},
  {Chambers}, {Chun}, {Heasley}, {Hodapp}, {Hunt}, {Jedicke}, {Jewitt},
  {Kudritzki}, {Luppino}, {Maberry}, {Magnier}, {Monet}, {Onaka}, {Pickles},
  {Rhoads}, {Simon}, {Szalay}, {Szapudi}, {Tholen}, {Tonry}, {Waterson}, \&
  {Wick}}]{Kaiser2002}
{Kaiser}, N., {Aussel}, H., {Burke}, B.~E., {et~al.} 2002, in Society of
  Photo-Optical Instrumentation Engineers (SPIE) Conference Series, Vol. 4836,
  Survey and Other Telescope Technologies and Discoveries, ed. J.~A. {Tyson} \&
  S.~{Wolff}, 154--164

\bibitem[{{Lemon} {et~al.}(2019){Lemon}, {Auger}, \& {McMahon}}]{lemon2019}
{Lemon}, C.~A., {Auger}, M.~W., \& {McMahon}, R.~G. 2019, \mnras, 483, 4242

\bibitem[{{LSST Dark Energy Science Collaboration} {et~al.}(2021){LSST Dark
  Energy Science Collaboration}, {Abolfathi}, {Armstrong}, {Awan}, {Babuji},
  {Bauer}, {Beckett}, {Biswas}, {Bogart}, {Boutigny}, {Chard}, {Chiang},
  {Cohen-Tanugi}, {Connolly}, {Daniel}, {Digel}, {Drlica-Wagner}, {Dubois},
  {Gawiser}, {Glanzman}, {Habib}, {Hearin}, {Heitmann}, {Hernandez},
  {Hlo{\v{z}}ek}, {Hollowed}, {Jarvis}, {Jha}, {Bryce Kalmbach}, {Kelly},
  {Kovacs}, {Korytov}, {Krughoff}, {Lage}, {Lanusse}, {Larsen}, {Li},
  {Longley}, {Lupton}, {Mandelbaum}, {Mao}, {Marshall}, {Meyers}, {Park},
  {Peloton}, {Perrefort}, {Perry}, {Plaszczynski}, {Pope}, {Rykoff},
  {S{\'a}nchez}, {Schmidt}, {Uram}, {Villarreal}, {Walter}, {Wiesner}, \&
  {Wood-Vasey}}]{DESC2021}
{LSST Dark Energy Science Collaboration}, {Abolfathi}, B., {Armstrong}, R.,
  {et~al.} 2021, arXiv e-prints, arXiv:2101.04855

\bibitem[{{Lucy}(1994)}]{Lucy1994}
{Lucy}, L.~B. 1994, in The Restoration of HST Images and Spectra - II, ed.
  R.~J. {Hanisch} \& R.~L. {White}, 79

\bibitem[{{MacLeod} {et~al.}(2010){MacLeod}, {Ivezi{\'c}}, {Kochanek},
  {Koz{\l}owski}, {Kelly}, {Bullock}, {Kimball}, {Sesar}, {Westman}, {Brooks},
  {Gibson}, {Becker}, \& {de Vries}}]{MacLeod2010}
{MacLeod}, C.~L., {Ivezi{\'c}}, {\v{Z}}., {Kochanek}, C.~S., {et~al.} 2010,
  \apj, 721, 1014

\bibitem[{{Magain} {et~al.}(1998){Magain}, {Courbin}, \& {Sohy}}]{Magain1998}
{Magain}, P., {Courbin}, F., \& {Sohy}, S. 1998, \apj, 494, 472

\bibitem[{{Michalewicz} {et~al.}(2023){Michalewicz}, {Millon}, {Dux}, \&
  {Courbin}}]{Michalewicz2023}
{Michalewicz}, K., {Millon}, M., {Dux}, F., \& {Courbin}, F. 2023, The Journal
  of Open Source Software, 8, 5340

\bibitem[{{Millon} {et~al.}(2020){Millon}, {Courbin}, {Bonvin}, {Paic},
  {Meylan}, {Tewes}, {Sluse}, {Magain}, {Chan}, {Galan}, {Joseph}, {Lemon},
  {Tihhonova}, {Anderson}, {Marmier}, {Chazelas}, {Lendl}, {Triaud}, \&
  {Wyttenbach}}]{Millon2020}
{Millon}, M., {Courbin}, F., {Bonvin}, V., {et~al.} 2020, \aap, 640, A105

\bibitem[{{Oguri} \& {Marshall}(2010)}]{Oguri2010}
{Oguri}, M. \& {Marshall}, P.~J. 2010, \mnras, 405, 2579

\bibitem[{{Refsdal}(1964)}]{Refsdal1964}
{Refsdal}, S. 1964, \mnras, 128, 307

\bibitem[{{Saha} {et~al.}(2024){Saha}, {Sluse}, {Wagner}, \&
  {Williams}}]{Saha2024}
{Saha}, P., {Sluse}, D., {Wagner}, J., \& {Williams}, L. L.~R. 2024, \ssr, 220,
  12

\bibitem[{{Schneider} {et~al.}(1992){Schneider}, {Ehlers}, \&
  {Falco}}]{Schneider1992}
{Schneider}, P., {Ehlers}, J., \& {Falco}, E.~E. 1992, {Gravitational Lenses}

\bibitem[{{Seitz} \& {Schneider}(1997)}]{SeitzSchneider1997}
{Seitz}, C. \& {Schneider}, P. 1997, \aap, 318, 687

\bibitem[{{Sersic}(1968)}]{Sersic1968}
{Sersic}, J.~L. 1968, {Atlas de Galaxias Australes}

\bibitem[{{Stetson}(1987)}]{Stetson1987}
{Stetson}, P.~B. 1987, \pasp, 99, 191

\bibitem[{{Suberlak} {et~al.}(2021){Suberlak}, {Ivezi{\'c}}, \&
  {MacLeod}}]{Suberlak2021}
{Suberlak}, K.~L., {Ivezi{\'c}}, {\v{Z}}., \& {MacLeod}, C. 2021, \apj, 907, 96

\bibitem[{{Sureau} {et~al.}(2020){Sureau}, {Lechat}, \& {Starck}}]{Sureau2020}
{Sureau}, F., {Lechat}, A., \& {Starck}, J.~L. 2020, \aap, 641, A67

\bibitem[{{Taak} \& {Treu}(2023)}]{Taak2023}
{Taak}, Y.~C. \& {Treu}, T. 2023, \mnras, 524, 5446

\bibitem[{{Tewes} {et~al.}(2013){Tewes}, {Courbin}, \& {Meylan}}]{Tewes2013}
{Tewes}, M., {Courbin}, F., \& {Meylan}, G. 2013, \aap, 553, A120

\bibitem[{{Trujillo} {et~al.}(2001){Trujillo}, {Aguerri}, {Cepa}, \&
  {Guti{\'e}rrez}}]{Trujillo2001}
{Trujillo}, I., {Aguerri}, J.~A.~L., {Cepa}, J., \& {Guti{\'e}rrez}, C.~M.
  2001, \mnras, 328, 977

\bibitem[{{Udalski}(2003)}]{Udalski2003}
{Udalski}, A. 2003, \actaa, 53, 291

\bibitem[{{Vegetti} {et~al.}(2023){Vegetti}, {Birrer}, {Despali}, {Fassnacht},
  {Gilman}, {Hezaveh}, {Perreault Levasseur}, {McKean}, {Powell}, {O'Riordan},
  \& {Vernardos}}]{Vegetti2023}
{Vegetti}, S., {Birrer}, S., {Despali}, G., {et~al.} 2023, arXiv e-prints,
  arXiv:2306.11781

\bibitem[{{Verde} {et~al.}(2023){Verde}, {Sch{\"o}neberg}, \&
  {Gil-Mar{\'\i}n}}]{Verde2023}
{Verde}, L., {Sch{\"o}neberg}, N., \& {Gil-Mar{\'\i}n}, H. 2023, arXiv
  e-prints, arXiv:2311.13305

\bibitem[{{Vernardos}(2022)}]{Vernardos2022}
{Vernardos}, G. 2022, \mnras, 511, 4417

\bibitem[{{Wong} {et~al.}(2020){Wong}, {Suyu}, {Chen}, {Rusu}, {Millon},
  {Sluse}, {Bonvin}, {Fassnacht}, {Taubenberger}, {Auger}, {Birrer}, {Chan},
  {Courbin}, {Hilbert}, {Tihhonova}, {Treu}, {Agnello}, {Ding}, {Jee},
  {Komatsu}, {Shajib}, {Sonnenfeld}, {Blandford}, {Koopmans}, {Marshall}, \&
  {Meylan}}]{Wong2020}
{Wong}, K.~C., {Suyu}, S.~H., {Chen}, G. C.~F., {et~al.} 2020, \mnras, 498,
  1420

\bibitem[{{Zhang} {et~al.}(2023){Zhang}, {Almoubayyed}, {Mandelbaum}, {Meyers},
  {Jarvis}, {Kannawadi}, {Schmitz}, {Guinot}, \& {LSST Dark Energy Science
  Collaboration}}]{Zhang2023}
{Zhang}, T., {Almoubayyed}, H., {Mandelbaum}, R., {et~al.} 2023, \mnras, 520,
  2328

\bibitem[{{Zuntz} {et~al.}(2018){Zuntz}, {Sheldon}, {Samuroff}, {Troxel},
  {Jarvis}, {MacCrann}, {Gruen}, {Prat}, {S{\'a}nchez}, {Choi}, {Bridle},
  {Bernstein}, {Dodelson}, {Drlica-Wagner}, {Fang}, {Gruendl}, {Hoyle}, {Huff},
  {Jain}, {Kirk}, {Kacprzak}, {Krawiec}, {Plazas}, {Rollins}, {Rykoff},
  {Sevilla-Noarbe}, {Soergel}, {Varga}, {Abbott}, {Abdalla}, {Allam}, {Annis},
  {Bechtol}, {Benoit-L{\'e}vy}, {Bertin}, {Buckley-Geer}, {Burke}, {Carnero
  Rosell}, {Carrasco Kind}, {Carretero}, {Castander}, {Crocce}, {Cunha},
  {D'Andrea}, {da Costa}, {Davis}, {Desai}, {Diehl}, {Dietrich}, {Doel},
  {Eifler}, {Estrada}, {Evrard}, {Fausti Neto}, {Fernandez}, {Flaugher},
  {Fosalba}, {Frieman}, {Garc{\'\i}a-Bellido}, {Gaztanaga}, {Gerdes},
  {Giannantonio}, {Gschwend}, {Gutierrez}, {Hartley}, {Honscheid}, {James},
  {Jeltema}, {Johnson}, {Johnson}, {Kuehn}, {Kuhlmann}, {Kuropatkin}, {Lahav},
  {Li}, {Lima}, {Maia}, {March}, {Martini}, {Melchior}, {Menanteau}, {Miller},
  {Miquel}, {Mohr}, {Neilsen}, {Nichol}, {Ogando}, {Roe}, {Romer}, {Roodman},
  {Sanchez}, {Scarpine}, {Schindler}, {Schubnell}, {Smith}, {Smith},
  {Soares-Santos}, {Sobreira}, {Suchyta}, {Swanson}, {Tarle}, {Thomas},
  {Tucker}, {Vikram}, {Walker}, {Wechsler}, {Zhang}, \& {DES
  Collaboration}}]{Zuntz2018}
{Zuntz}, J., {Sheldon}, E., {Samuroff}, S., {et~al.} 2018, \mnras, 481, 1149

\end{thebibliography}

\end{document}